\documentclass[prx,aps,showpacs,superscriptaddress,nofootinbib]{revtex4-1}
\usepackage[top=3cm, bottom=2cm, left=2cm, right=2cm]{geometry}
\usepackage{graphicx}
\usepackage{amsbsy}
\usepackage{amsmath}
\usepackage{xcolor}
\usepackage{amssymb}

\makeatletter
\@booleanfalse\titlepage@sw
\makeatother
\begin{document}
\title{Many-body localization in waveguide QED}

\author{N. Fayard}
\address{ICFO - Institut de Ci{\`e}ncies Fot{\`o}niques, The Barcelona Institute of Science and Technology, 08860 Castelldefels, Barcelona, Spain}
\author{L. Henriet}
\address{Pasqal, 2 avenue Augustin Fresnel, 91120 Palaiseau, France}
\author{A. Asenjo-Garcia}
\address{Department of Physics, Columbia University, New York, New York 10027, USA}
\author{D.E. Chang}
\address{ICFO - Institut de Ci{\`e}ncies Fot{\`o}niques, The Barcelona Institute of Science and Technology, 08860 Castelldefels, Barcelona, Spain}
\address{ICREA - Instituci{\'o} Catalana de Recerca i Estudis Avan{\c c}ats, 08015 Barcelona, Spain}


\date{\today}
\begin{abstract}
At the quantum many-body level, atom-light interfaces generally remain challenging to solve for or understand in a non-perturbative fashion. Here, we consider a waveguide quantum electrodynamics model, where two-level atoms interact with and via propagating photons in a one-dimensional waveguide, and specifically investigate the interplay of atomic position disorder, multiple scattering of light, quantum nonlinear interactions and dissipation. We develop qualitative arguments and present numerical evidence that such a system exhibits a many-body localized~(MBL) phase, provided that atoms are less than half excited. Interestingly, while MBL is usually formulated with respect to closed systems, this system is intrinsically open. However, as dissipation originates from transport of energy to the system boundaries and the subsequent radiative loss, the lack of transport in the MBL phase makes the waveguide QED system look essentially closed and makes applicable the notions of MBL. Conversely, we show that if the system is initially in a delocalized phase due to a large excitation density, rapid initial dissipation can leave the system unable to efficiently transport energy at later times, resulting in a dynamical transition to an MBL phase. These phenomena can be feasibly realized in state-of-the-art experimental setups. 
\end{abstract}
\maketitle

\section{Introduction}
Quantum light-matter interfaces are being actively investigated, for their many possibilities to explore fundamental science and for applications. However, the full quantum many-body problem of interacting atoms and photons remains highly challenging to understand, not only due to the large Hilbert space, but also its out-of-equilibrium and open nature. Most of our theoretical knowledge is thus restricted to certain regimes. For example, much of our quantum theory, such as to model applications, treats the atomic medium as being a smooth quantum field~\cite{hammerer2010quantum}, thus ignoring the potential complexities associated with granularity, disorder averaging, and multiple scattering of light. On the other hand, there has been significant work to deal with granularity and multiple scattering~\cite{fleischhauer1999radiative,javanainen2016light,javanainen2014shifts,bromley2016collective,cottier2018role,schilder2020near,skipetrov2014absence,pellegrino2014observation}, but largely restricted to classical or mean-field limits, where potentially strong non-linear optical interactions and the buildup of quantum correlations are ignored.

It is hence important to more fully develop insights into what opportunities and many-body quantum phenomena might arise from the combination of multiple scattering and nonlinear interactions between atoms and light~\cite{masson2020many,poddubny2020quasiflat, zhang2020subradiant, zhong2020classification,williamson2020superatom}. For example, it has been suggested that multiple scattering can be enhanced using ordered atomic arrays and become an important resource for applications, which can lead to spin squeezing interactions~\cite{qu2019spin}, improved photon-photon gates~\cite{moreno2021quantum,zhang2021photon}, or prolonged interrogation times for optical lattice clocks through the suppression of spontaneous emission~\cite{henriet_clock,ostermann2013protected}. Separately, in the regime of dilute, disordered media, perturbative diagrammatic techniques to account for nonlinear interactions and multiple scattering have only recently been developed~\cite{wellens}. Aside from being a possible resource, the effects of multiple scattering could also impose new, unexpected limits on the performance of disordered quantum light-matter interfaces~\cite{he2021unraveling,Andreoli2021}.

Here, we propose the existence of and theoretically investigate a non-perturbative, quantum many-body phenomenon in a minimal ``waveguide quantum electrodynamics (QED)'' model, consisting of disordered two-level atoms or qubits interacting with photons in an infinite one-dimensional (1D) waveguide. In particular, we provide qualitative arguments and quantitative numerical simulations to argue that the system can support a many-body localized phase. The topic of many-body localization~(MBL) has generated intense interest recently within condensed matter physics~\cite{basko,mirlin,nandkishore2015many,alet2018many,abanin2019colloquium}. A simple phenomenological Hamiltonian is hypothesized to capture the novel MBL phase~\cite{huse2014phenomenology,serbyn2013local}, which allows for predictions of exotic many-body properties such as the failure of a system to thermalize, the absence of transport, and the slow growth of entanglement entropy following a quench. Interestingly, while the concept of MBL is usually formulated with respect to closed systems, our waveguide QED system is intrinsically open, as light emitted by any finite set of excited atoms will be irreversibly lost once passing the boundaries of the atomic ensemble. This results in an additional rich interplay between dissipation and essentially closed, MBL-like dynamics. In particular, we show that an MBL phase occurs at low atomic excitation densities (approximately less than half excited in the large atom limit). The absence of transport then suppresses propagation of energy to the system boundaries and the resulting dissipation, which in turn makes the notion of MBL applicable in the first place. On the other hand, for systems that are initially highly excited, transport can be responsible for a rapid dissipation at early times. This causes the atomic excitation density to drop until transport is no longer efficient, resulting in a \textit{dynamical transition} into an MBL phase. This physics should be feasible to explore in existing state-of-the-art waveguide QED systems, such as superconducting quantum bits coupled to microwave transmission lines~\cite{wallraff,hoi,mirhosseini2019cavity} or structured waveguides~\cite{houck,ferreira2020collapse}.

The rest of the paper is organized as follows. First, in Sec.~\ref{sec:description}, we introduce the waveguide QED Hamiltonian and present a qualitative argument that an MBL phase should exist in the thermodynamic limit, provided that the atoms are less than half excited. Given the large Hilbert space of the system, consisting of the exponentially large space associated with the atoms and the infinite space associated with the continuum of photon modes, we then restrict ourselves to the regime of near-resonant photon interactions. Then, in Sec.~\ref{sec:spinmodel}, we show how the photonic modes can be integrated out, reducing the system to an open, interacting ``spin model'' describing photon-mediated interactions between the atomic degrees of freedom. Before going to MBL, and to gain intuition, we show how Anderson localization manifests itself in the spin model formalism, within the linear optics~(single-excitation) limit. This example illustrates that although the system is formally open, localization implies that dynamics of a large system is well-described just by the Hermitian Hamiltonian component of the spin model. We briefly introduce the phenomenology of MBL phases in Sec.~\ref{sec:phenomenologyMBL}. As MBL is typically studied with respect to closed systems, and given the observation above regarding the connection between localization and Hermitianity, in Sec.~\ref{sec:MBLclosed} we present numerical evidence of MBL for the Hermitian Hamiltonian, namely the absence of transport and logarithmic growth of entanglement entropy for sufficiently low excitation densities. In Sec.~\ref{sec:MBLopen} we then consider the physical open system, and propose and numerically investigate a number of realistic observables for MBL. At low excitation densities, we again find an MBL phase. Surprisingly, however, for higher excitation densities, where MBL is not observed for the closed system, we find that transport to the system boundaries and the corresponding dissipation enable the system to initially and rapidly lower the excitation density, and thus dynamically transition into an MBL phase. Finally, in Sec.~\ref{sec:Conclusion}, we provide an outlook of possible future interesting directions to explore, and discuss the prospects of experimentally observing such physics in state-of-the-art waveguide QED platforms.

\section{Description of the system}\label{sec:description}

\begin{figure}[t]   
\begin{center}
     \includegraphics[width=0.80\textwidth]{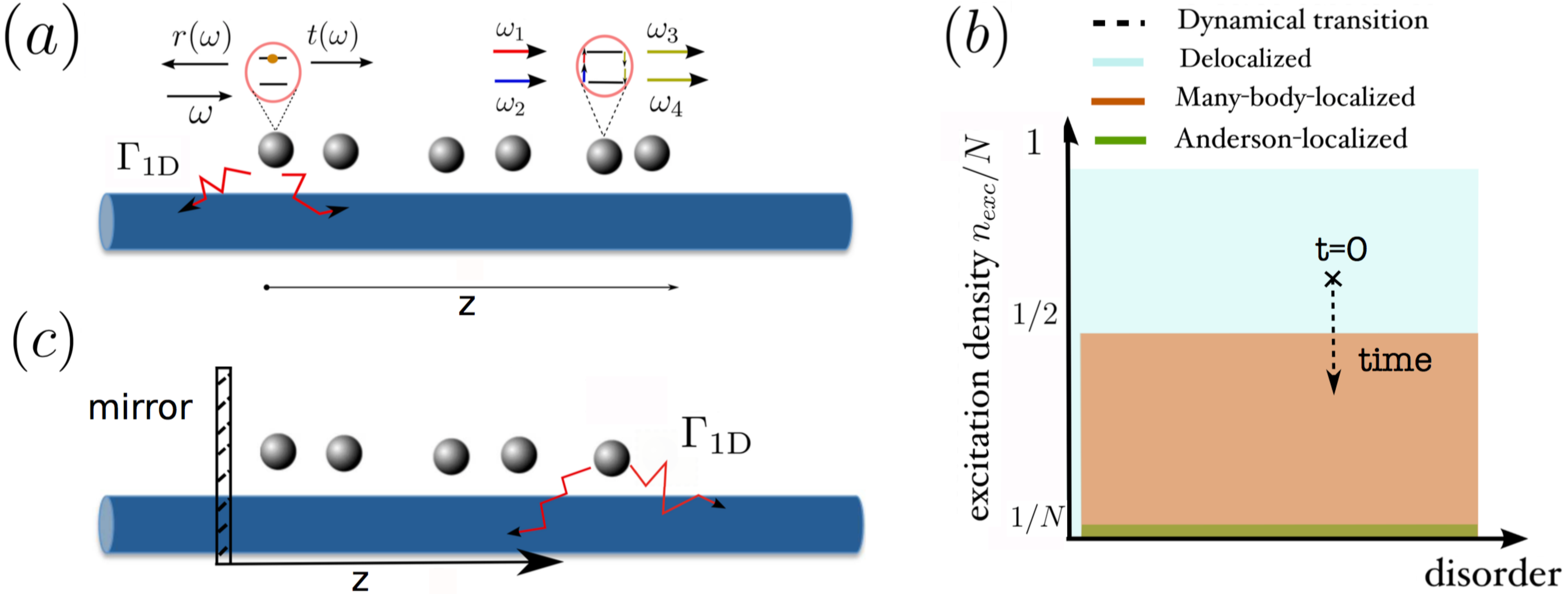} 
             \end{center}
     \caption{ (a) Schematic illustration of a spatially disordered atomic chain coupled to photons in a one-dimensional, bi-directional waveguide. The magnitude of  atom-light interactions is set by the rate $\Gamma_\text{1D}$ by which a single, isolated excited atom irreversibly emits a photon into the waveguide. With multiple atoms and excitations, the system dynamics becomes complex and interacting due to the following possibilities. First, a photon can be reflected or transmitted with non-trivial, frequency-dependent amplitudes $r(\omega),t(\omega)$ upon scattering off of an atom, which then leads to multiple scattering in the presence of multiple atoms. Second, multiple photons (e.g., two photons with frequencies $\omega_{1,2}$ illustrated) scattering off of the same atom can lead to the generation of entangled, frequency-mixed photons ($\omega_3+\omega_4=\omega_1+\omega_2$), which also become multiply scattered. (b) Proposed phase diagram in the thermodynamic limit of large atom number $N\rightarrow\infty$, versus amount of disorder and fraction of atoms excited, $n_{\rm exc}/N$. For a single excitation, $n_{\rm exc}/N=1/N$, the system is non-interacting, and arbitrarily small disorder gives rise to Anderson localization. Beyond a single excitation, the system becomes interacting due to the nonlinearity of two-level atoms. Up to $n_{\rm exc}/N=1/2$, we hypothesize that the system exhibits MBL. For $n_{\rm exc}/N>1/2$, photon transport is allowed and leads to delocalization. Going beyond the thermodynamic limit, and considering a finite atomic system, transport in the delocalized phase leads to dissipation, in the form of emission of light beyond the atomic system boundaries. The subsequent loss of excitation density can lead to a dynamical transition into the MBL phase, as depicted by the dashed arrow showing time evolution. (c) Schematic illustration of a spatially disordered atomic chain coupled to photons in a one-dimensional, ``half'' waveguide, with one end closed by a mirror at $z=0$. The atomic system can thus only dissipate by emitting photons from one side of the chain.}\label{fig:schematic} 
\end{figure}

Our system of interest is represented in Fig.~\ref{fig:schematic}(a). It consists of a one-dimensional, bi-directional waveguide, whose photons couple to the ground-excited ($|g\rangle$-$|e\rangle$) transitions of $N$ two-level atoms located at random positions $z_i$. The atomic transition frequency is given by $\omega_0$. A minimal model for such a system is given by the following Hamiltonian (with $\hbar=1$):
 \begin{equation}
\mathcal{H}=\underbrace{\omega_0\sum_{i}\hat{\sigma}_{ee}^i}_{\mathcal{H}_{\rm a}} +\underbrace{\sum_{\nu=\pm}\int \mathrm dk\;\omega_k \hat{b}^{\dagger}_{\nu,k}\hat{b}_{\nu,k}}_{\mathcal{H}_{\rm ph}} +\underbrace{g\sum_{\nu,i}\int\mathrm dk\left(\hat{b}_{\nu,k}\hat{\sigma}_{eg}^i e^{i\nu kz_i} + \operatorname{h.c.} \right)}_{\mathcal{H}_{\rm int}}. \label{eq:Hfull}
\end{equation}
Here, $\mathcal{H}_{\rm a}$ describes the excited-state energy of the atoms, and we adopt the notation $\hat{\sigma}_{\alpha\beta}^i=|\alpha_i\rangle\langle\beta_i|$ with $\{\alpha,\beta\}\in\{g,e\}$ for the operators of atom $i$. $\mathcal{H}_{\rm ph}$ describes the energy of the photons, characterized by a continuum of modes with two possible propagation directions (labeled $\nu=\pm$) and wavevector $k$. We assume that within the bandwidth of modes to which the atoms significantly couple, the dispersion relation for the guided
modes can be linearized as $\omega_k=c\vert k \vert$. The interactions, as given by $\mathcal{H}_{\rm int}$, describe the processes by which an excited atom can emit a photon, or a ground-state atom can absorb a photon, with a coupling constant $g$ assumed to be identical for all atoms, and an interaction phase $e^{i\nu kz_i}$ encoding the photon propagation. In the following, we consider disorder in the positions of the atoms: $z_i=(i+\epsilon_i)d$, with $d$ the average distance between two neighboring atoms and $\epsilon_i$ a random variable. For numerics, we will focus on the regime of full disorder, where $\epsilon_i$ is drawn between $-1/2$ and $1/2$, in order to minimize the effective localization length of the system and the number of atoms needed in  simulations, although we believe our conclusions are general for any amount of disorder (see below). The Hamiltonian of Eq.~(\ref{eq:Hfull}) can be an excellent approximation for a number of systems such as superconducting qubits coupled to a transmission line~\cite{wallraff,hoi,mirhosseini2019cavity}, or atoms coupled to a photonic crystal waveguide~\cite{goban2015superradiance,hood2016atom}, where additional sources of dissipation (not included in this model) can potentially be much smaller than the coherent atom-waveguide interactions.

To gain some basic insight into Eq.~(\ref{eq:Hfull}), we begin by relating it to well-known results regarding Anderson localization of light in 1D, in the linear optical or single-excitation limit with disorder in the position of the atoms.
The single-photon, single-atom scattering dynamics of Eq.~(\ref{eq:Hfull}) can be exactly solved~\cite{shen2005coherent,chang2007single}. In particular, the response of a ground-state atom to an incoming photon of well-defined frequency $\omega$ is characterized by the reflection and transmission coefficients $r(\omega)=-\Gamma_\text{1D}/(\Gamma_\text{1D}-2i\Delta)$ and $t(\omega)=1+r(\omega)$~(Fig.~\ref{fig:schematic}(a), with $\Delta=\omega-\omega_0$ being the detuning between the photon and the atomic transition, and $\Gamma_\text{1D}=4\pi g^2/c$ being the spontaneous emission rate of a single, isolated excited atom into the waveguide. When many atoms are coupled together via the waveguide, the multiple scattering arising from the succession of transmission and reflection events~(Fig.~\ref{fig:schematic}(a) by different atoms can be solved by the transfer matrix formalism~\cite{muller2010disorder}. Performing a disorder average over the position fluctuations $\epsilon_i$ of the N atoms, the 1D waveguide is always Anderson localized for arbitrarily small disorder (arbitrarily narrow distributions of $\epsilon_i$, see Fig.~\ref{fig:schematic}(b)~\cite{anderson1958absence,abrahams1979scaling,billy2008direct}. The transmittance exhibits exponential attenuation with large fluctuations $\overline{\operatorname{log}(T_{\rm tot})}=-N/N_{\rm loc}$ and $\operatorname{var}(\operatorname{log}(T_{\rm tot}))=2N/N_{\rm loc}$, with $\overline{(...)}$ denoting the average value over the different disorder realizations and $\operatorname{var}(...)$ the variance of the random variable. $N_{\rm loc}$ is the localization length (expressed in terms of number of atoms), and is related to the single-atom linear transmittance $T(\Delta)=|t(\Delta)|^2$ by $N_{\rm loc}(\Delta)= 1/|\operatorname{log} T(\Delta)| = |\operatorname{log} (4\Delta^2/(\Gamma_\text{1D}^2+4\Delta^2))|^{-1}$ in the dilute regime~($d>\lambda=2\pi c/\omega$) and for large disorder~\cite{muller2010disorder}. This quantity goes to zero at resonance, and increases with increasing detuning $|\Delta|$ as the atomic response becomes weaker away from resonance. 
For systems larger than the localization length $N>N_{\rm loc}$, strong destructive interference suppresses the propagation of light through the medium.

While rigorously establishing an MBL transition beyond the single-excitation limit is more complicated, we can nonetheless make a qualitative argument, by calculating a nonlinear single-atom transmittance $T(\Delta,\rho_{ee})$ that depends on the excited state population $\rho_{ee}$, and then substituting into the Anderson localization result $N_{\rm loc}(\Delta,\rho_{ee}) \sim 1/|\operatorname{log} T(\Delta,\rho_{ee})|$ to estimate a length scale over which transport becomes prohibited. As discussed in the Appendix, this transmittance can be exactly calculated for an incident coherent state field of arbitrary amplitude, with corresponding Rabi frequency $\Omega$. One finds the relations  $\rho_{ee}=\frac{\Omega^2}{\Gamma_\text{1D}^2+4\Delta^2+2\Omega^2}$ and  $T=\frac{4\Delta^2+8\Omega^2}{\Gamma_\text{1D}^2+4\Delta^2+8\Omega^2}$. Notably, for large photon input ($\Omega\rightarrow\infty$), one finds that $\rho_{ee}\rightarrow 1/2$ and $T\rightarrow 1$. This expresses the well-known result that an atom becomes saturated at very high intensities and is no longer able to respond to light. In turn, reflection and multiple scattering are suppressed, and the localization length $N_{\rm loc}(\Delta,\rho_{ee}\rightarrow 1/2)\rightarrow \infty$ diverges. We thus hypothesize that an atomic excitation density of $n_{\rm exc}/N=1/2$ sets the MBL-delocalization transition, as illustrated in Fig \ref{fig:schematic}(b).  

\section{Spin model}\label{sec:spinmodel}

Beyond a single excitation, directly solving Eq.~(\ref{eq:Hfull}) is difficult due to the infinite Hilbert space of the photons and frequency mixing induced by the atoms. Furthermore, the localization length $N_{\rm loc}(\Delta,\rho_{ee})$ increases both with increasing detuning and atomic population, and can rapidly exceed numerically tractable system sizes (see Appendix).

To partially mitigate these challenges, we will restrict ourselves to the regime in which the photons involved in the dynamics are near resonance with the atoms. As discussed in Ref.~\cite{Chang_2012}, since the field part of the Hamiltonian~(\ref{eq:Hfull}) is quadratic, it can be formally integrated out, resulting in photon-mediated interactions between atoms. Furthermore, near resonance, the delay time of these interactions due to the speed of light can be ignored to good approximation. This is because the atoms in typical situations have very large ratios of resonance frequencies to linewidths, $\omega_0\gg \Gamma_\text{1D}$. Thus, the atoms are very dispersive, for example, as characterized by the large, frequency-dependent phase shifts in $r(\omega),t(\omega)$, and this causes the propagation delay of near-resonant photons to be dominated by interaction with the atoms, rather than the speed of light itself. Equivalently, the hybrid atom-photon polaritons that diagonalize the system are in fact almost entirely atomic in nature. Ignoring retardation, one obtains a reduced master equation describing instantaneous photon-mediated interactions, between only the atomic ``spin'' degrees of freedom~\cite{Chang_2012,lalumiere2013input,Caneva_2015,asenjo2017atom,albrecht2019subradiant},
\begin{equation}
\dot{\hat{\rho}}=-i \left[ \mathcal{H}_\text{1D}\hat{\rho}-\hat{\rho} \mathcal{H}_\text{1D}^{\dagger} \right] + \sum_{i,j}\Gamma_{i,j}\hat{\sigma}_{ge}^i\hat{\rho}\hat{\sigma}_{eg}^j, \label{eq:master}
\end{equation}
with
\begin{equation}
\mathcal{H}_\text{1D}=-i\frac{\Gamma_\text{1D}}{2}\sum_{i,j=1}^N\operatorname{exp}(-ik_\text{1D}\vert z_i-z_j\vert)\hat{\sigma}_{eg}^i\hat{\sigma}_{ge}^j
\end{equation}
 and  $\Gamma_{i,j}=\Gamma_\text{1D}\operatorname{cos}(k_\text{1D}\vert z_i-z_j\vert)$ and $k_\text{1D}=\omega_0/c$. Here, we have transformed to a rotating frame, so that the trivial phase evolution associated with $\mathcal{H}_{\rm a}$ can be ignored. We note that in one dimension, the photon-mediated interactions between atoms are infinite in range. Furthermore, integrating out the photons results in a dissipative (open) system, as atoms can physically emit a photon that goes beyond the atomic system boundaries, resulting in irreversible loss of excited population.

If loss is indeed associated with the system boundaries, then conceptually its effects (for a fixed initial number of excitations) can be reduced by considering progressively larger system sizes, where the ratio of ``boundary''~(\textit{e.g.}, atoms within a distance $\sim N_{\rm loc}$ of an edge) to ``bulk'' regions decreases. While the maximum number of atoms in numerics is constrained, we can nonetheless decrease this ratio by introducing a closely related ``half-waveguide'' model, as illustrated in Fig.~\ref{fig:schematic}(c). Here, one end of the waveguide (say at $z=0$) is terminated by a perfect mirror, which results in only one open boundary where photons can escape. Applying similar considerations as the derivation of Eq.~(2), it is straightforward to show that the master equation is
\begin{equation}
\dot{\hat{\rho}}=-i \left[ \mathcal{H}_{\rm half}\hat{\rho}-\hat{\rho} \mathcal{H}_{\rm half}^{\dagger} \right] + \sum_{i,j}\Gamma_{i,j}\hat{\sigma}_{ge}^i\hat{\rho}\hat{\sigma}_{eg}^j, \label{eq:masterhalf}
\end{equation}
with
\begin{equation}
\mathcal{H}_{\rm half}=-i\Gamma_\text{1D}/2\sum_{i,j=1}^N\left[\operatorname{exp}(-ik_\text{1D}\vert z_i-z_j\vert )-\operatorname{exp}(-ik_\text{1D}\vert z_i+z_j\vert ) \right]\hat{\sigma}_{eg}^i\hat{\sigma}_{ge}^j
\label{eq:Hhalf}
\end{equation}
and  $\Gamma_{i,j}=\Gamma_\text{1D}\left[\operatorname{cos}(k_\text{1D}\vert z_j-z_i\vert)-\operatorname{cos}(k_\text{1D}\vert z_j+z_i\vert) \right]$.

While Eqs.~(\ref{eq:master}) and~(\ref{eq:masterhalf}) have now eliminated the infinite Hilbert space of photons, an interesting new feature emerges: dissipation. In particular, MBL is usually formulated with respect to closed systems. Furthermore, in experimental platforms to investigate MBL thus far, any degree of dissipation is viewed as an unwanted imperfection on top of an idealized closed system. Here, in contrast, the dissipative part of the master equation a priori has the same strength~($\sim \Gamma_\text{1D}$) as the coherent interactions. A key idea that we aim to establish is that within an MBL regime, the bulk region of a many-body system behaves as if it is increasingly closed as it becomes further in distance from a boundary, as suppressed transport makes the emission of a photon past the system edges improbable. These ideas can be better clarified by returning to the simpler problem of Anderson localization, but now studied from the spin model perspective.

\subsection{Disorder in the single-photon limit}

\begin{figure}[t]   
\begin{center}
     \includegraphics[width=0.70\textwidth]{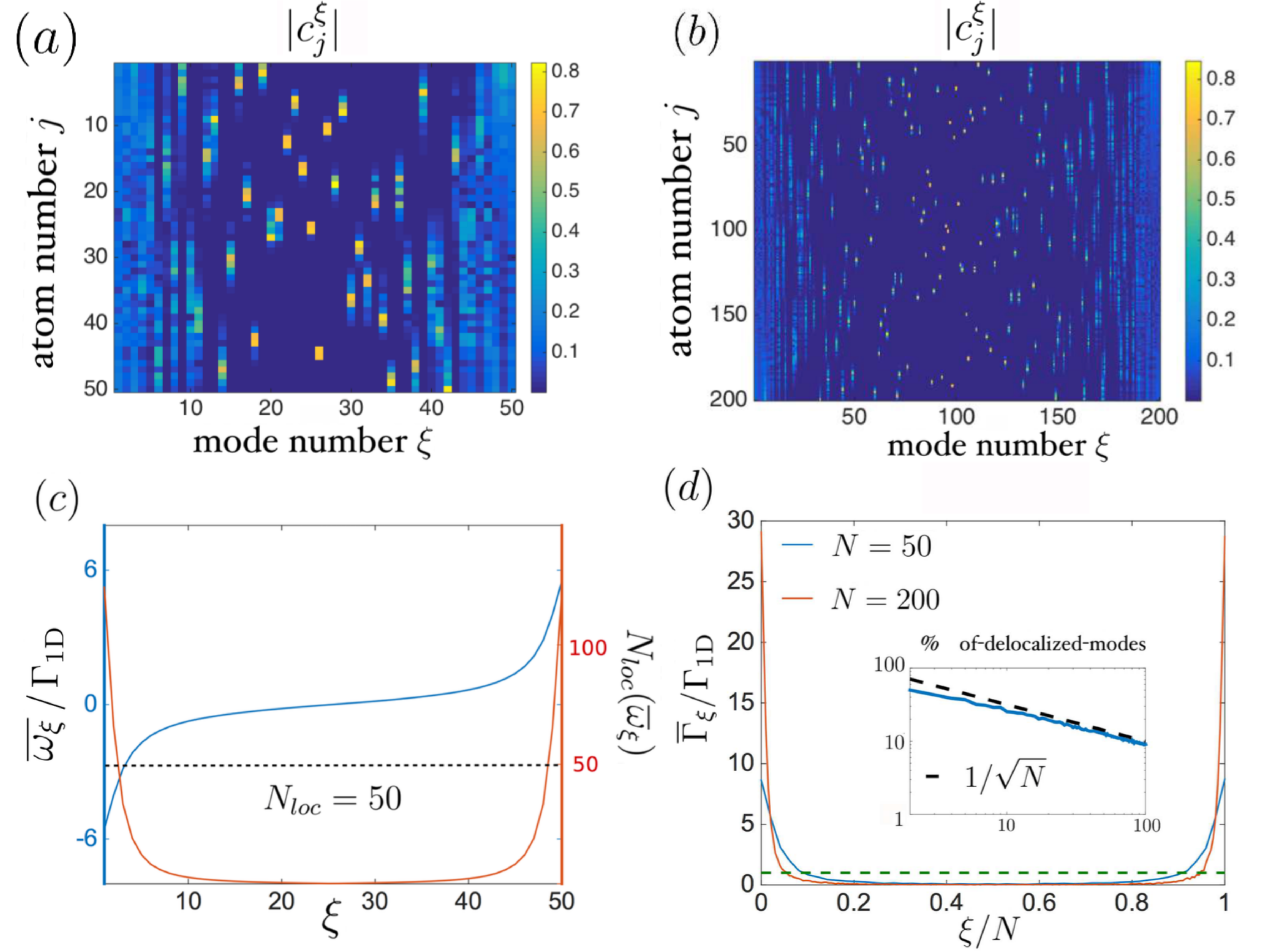}  
             \end{center}
     \caption{Spatial profile of the single-excitation eigenmodes of the effective Hamiltonian $\mathcal{H}_\text{1D}$, ordered with respect to the real part of their eigenvalue, for two different system sizes: $N=50$ (a) and $N=200$~(b). These eigenmodes have been calculated for a single disorder realization, and are indexed from lowest~($\xi=1$) to highest~($\xi=N$) frequency. (c) Average value over a large number of disorder realizations of $\omega_{\xi}/\Gamma_\text{1D}$~(blue) as a function of $\xi$, together with the associated Anderson localization length $N_{\rm loc}(\overline{\omega}_{\xi})$~(orange) in units of number of atoms. This plot is calculated for $N=50$ atoms. For reference, we indicate where the localization length corresponds to the system size by the dashed black line. (d) Average value over a large number of disorder realizations of the normalized decay rate $\Gamma_\xi/\Gamma_\text{1D}$ of eigenmodes, as a function of $\xi/N$, for two different system sizes $N=50$~(blue) and $N=200$~(orange). A large fraction of modes have a decay rate smaller than $\Gamma_\text{1D}/2$~(dashed green line). Numerically, the percentage of modes with decay rate larger than $\Gamma_\text{1D}/2$  decreases approximately as $\sim 1/\sqrt{N}$~(inset).} \label{fig:collective} 
\end{figure} 

The key single-excitation properties are encoded in the $N$ single-excitation eigenmodes and eigenvalues of the non-Hermitian Hamiltonian $\mathcal{H}_\text{1D}$. The eigenvalues themselves are complex, with the real and imaginary parts accounting for the shift in resonance frequency $\omega_{\xi}$ of the collective mode with respect to the bare atomic frequency, and half of the collective decay rate $\Gamma_{\xi}/2$, respectively.
 
As a concrete example, we first take one single realization of disorder, and sort the eigenstates in increasing order of their resonance frequencies, labeled by $1\leq \xi \leq N$. Denoting the wave functions by $|\psi_{\xi}\rangle= \sum_j c_j^{\xi} \hat{\sigma}_{eg}^{j}|g\rangle^{\otimes N}$, in Figs.~\ref{fig:collective}(a,b) we plot $|c_j^{\xi}|$,  the square root of the probability for atom $j$ to be excited, for the atom numbers of $N=50$ and $N=200$, respectively. For all numerical calculations, we take an average atomic spacing of $d=2.7\pi/k_\text{1D}$~(to be in the dilute regime) and full disorder. We see that states in the middle of the spectrum ($\xi \sim N/2$) are localized, while states at the edges of the spectrum ($\xi \sim 1$ and $\xi \sim N$) are extended. Furthermore, in Fig.~\ref{fig:collective}(c), we fix the atom number $N=50$, and plot the average value of the resonance frequency $\overline{\omega}_{\xi}/\Gamma_\text{1D}$ versus $\xi$ over a large number of disorder realizations (blue curve). We simultaneously plot (orange curve) the previously established Anderson localization length $N_{\rm loc}(\overline{\omega}_{\xi})$ corresponding to these frequencies. Comparing with Fig.~\ref{fig:collective}(a), we see that the transition from localized to extended eigenstates occurs when $|\overline{\omega}_{\xi}|$ is detuned enough that the localization length $N_{\rm loc}(\overline{\omega}_{\xi})>N$ exceeds the number of atoms inside the chain. In particular, the existence of extended states of the spin model~(such as found in Refs.~\cite{haakh2016polaritonic, fedorovich2020disorder,goetschy2011non}) does not contradict the fact that Anderson localization always exists in the 1D disordered system. Specifically, for any detuning $\Delta$ there always exists some sufficiently large $N>N_{\rm loc}$ where one will \textit{eventually observe} Anderson localization.

In Fig.~\ref{fig:collective}(d), we plot the average value over a large number of disorder realizations of the decay rate $\overline{\Gamma}_{\xi}/\Gamma_\text{1D}$, normalized by the single-atom decay rate, as a function of $\xi/N$ for two different system sizes $N=50$ and $N=200$. We see that the extended modes in the edges of the spectrum have a large decay rate, while the localized modes in the middle have an exponentially small decay rate. This confirms that localized modes, with populations away from the boundaries, cannot efficiently decay by spontaneous emission. Furthermore, as $N$ increases, we find that the percentage of delocalized modes [here quantified by the percentage of modes with a decay rate greater than $\Gamma_\text{1D}/2$, above the green line in Fig.~\ref{fig:collective}(d)] decreases approximately as $\sim 1/\sqrt{N}$, as shown in the inset.

While this analysis was restricted to the single-excitation manifold, it naturally also follows that for our particular model, the existence of MBL cannot be inferred purely by looking for the localized nature of eigenstates of the spin model with large number of excitations. In particular, a study along those lines would have to determine whether an apparently delocalized state might become localized as the system size is increased. As that is difficult within numerical capabilities, we must demonstrate the existence of MBL using other measures. Separately, we note that while localized and extended single-excitation eigenstates can be identified by spectral filtering (i.e. looking at eigenstates at the center and edges of the spectrum, respectively), this procedure cannot be extended to more excitations. For example, assuming that interactions can be considered as a perturbation, two single-excitation extended eigenstates with energies $\pm \omega$ could combine to give an extended two-excitation eigenstate with approximately zero energy.

\subsection{Localization as a transition to Hermitianity}

\begin{figure}[b]   
\begin{center}
     \includegraphics[width=1\textwidth]{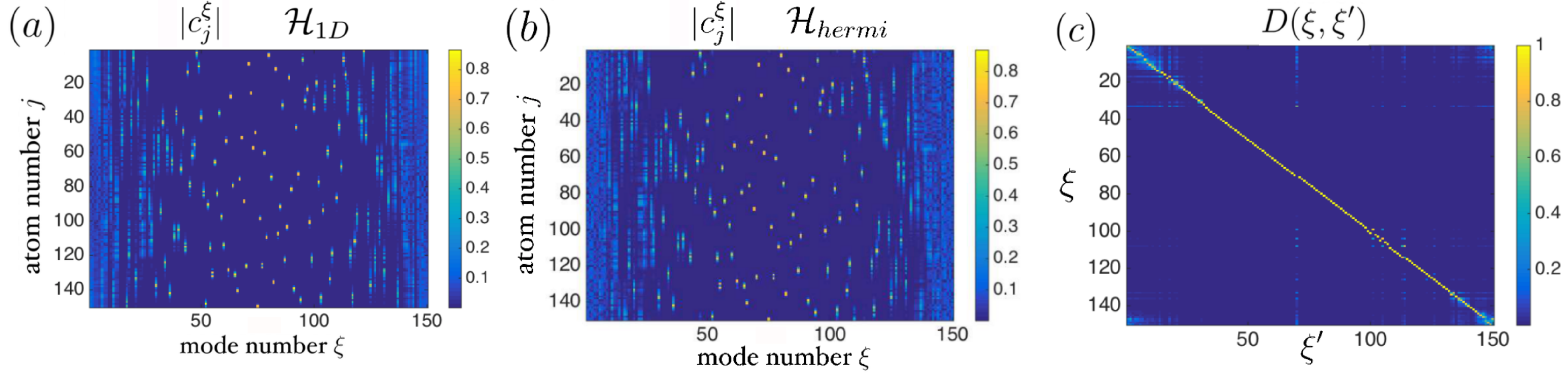} 
             \end{center}
   \caption{A representative plot of the excitation amplitude $|c_j^{\xi}|$ versus atom number $j$ and eigenmode index $\xi$, for a single disorder realization and within the single-excitation manifold. The excitation amplitudes are plotted for (a) the effective non-Hermitian Hamiltonian $\mathcal{H}_\text{1D}$ and (b) its Hermitian component $\mathcal{H}_{\rm hermi}$. (c) Magnitude $D(\xi,\xi')=|\langle\psi_{\xi}^\text{(1D)}|\psi_{\xi'}^{\rm (hermi)}\rangle|$ of the dot product between the eigenmodes of the two Hamiltonians. All of these plots were generated for a system of $N=150$ atoms. }\label{fig:hermitianity}  
\end{figure}

If Anderson (or many-body) localized states experience exponentially small decay rates versus distance of excitations from boundaries, the dissipative component of Eq.~(\ref{eq:master}) should have negligible effect, and such states should essentially be governed only by the Hermitian component $\mathcal{H}_{\rm hermi}=(\mathcal{H}_\text{1D}+\mathcal{H}_\text{1D}^{\dagger})/2$ of the effective Hamiltonian. This Hermitian component reads:
\begin{equation}
\mathcal{H}_{\rm hermi}=\frac{ \Gamma_\text{1D}}{2}\sum_{i,j} \operatorname{sin}\left(k_\text{1D}\vert z_i -z_j \vert\right) \hat{\sigma}_{ge}^i\hat{\sigma}_{eg}^j.
\label{eq:Hhermi}
\end{equation}

We can again easily confirm this in the single-excitation limit. Specifically, for a single disorder realization and $N=150$, we calculate the eigenstate amplitudes $\vert c_j^{\xi}\vert$ for the non-Hermitian and Hermitian Hamiltonians, which we plot in Figs.~\ref{fig:hermitianity}(a,b), respectively. Then, in Fig.~\ref{fig:hermitianity}(c), we plot the overlap $D(\xi,\xi')=|\langle\psi_{\xi}^\text{(1D)}|\psi_{\xi'}^{\rm (hermi)}\rangle|$  between the non-Hermitian and Hermitian eigenstates. We find that this matrix is nearly the identity, particularly in the middle of the spectrum where the modes are well-localized~\cite{hamazaki2019non}.

This suggests that we can establish the existence of an MBL in two complementary ways. First, we can take the Hermitian Hamiltonian $\mathcal{H}_{\rm hermi}$, and look for well-established mathematical signatures for closed systems, such as logarithmic growth of entanglement entropy following a quench. Then, assuming that the existence of MBL is not affected by losses for the reasons above, we can simultaneously look for realistic observables in the physical, \textit{open} system.

\section{Introduction to conventional MBL}\label{sec:phenomenologyMBL}

\begin{figure}[b]   
\begin{center}
     \includegraphics[width=0.90\textwidth]{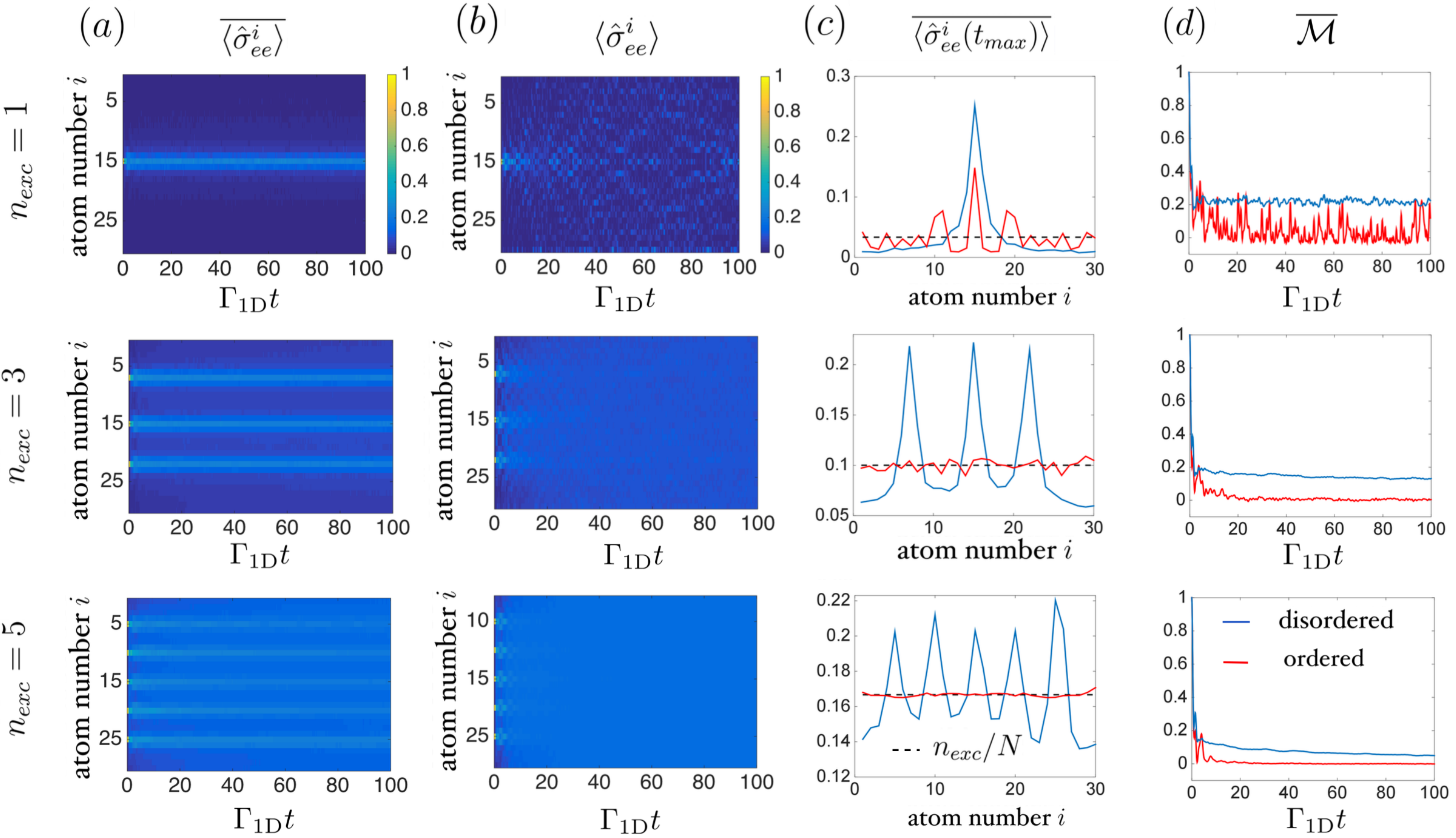} 
             \end{center}
   \caption{ Transport properties of $N=30$ atoms in the Hermitian waveguide, averaged over disorder configurations. Each line corresponds to a different number of excitations~($ n_{\rm exc}=1,3,5$), initialized in a product state. In (a) and (b), we plot the time-dependent, site-resolved excited state population $\langle \hat{\sigma}_{ee}^{i}(t) \rangle$, for disordered and ordered systems, respectively. (c) Site-resolved population $\langle \hat{\sigma}_{ee}^{i}(t_{\rm max}) \rangle$ evaluated at the final simulation time $t_{\rm max}=100/\Gamma_\text{1D}$, for both disordered and ordered systems. As a guide to the eye, we plot $n_{\rm exc}/N$~(dashed line), the population if the initial excitations were to become uniformly distributed. (d) Time-dependent memory parameter $\overline{\mathcal{M}}(t)$ for both disordered and ordered systems.}\label{fig:transportclosed}   
\end{figure}

The key properties of MBL can be understood from a ``canonical" Hamiltonian hypothesized for all MBL systems~\cite{serbyn2013local,huse2014phenomenology}. For spin systems like ours, it reads
\begin{equation}
\mathcal{H}_{\rm LIOM}=\sum_i \omega_i\hat{\tau}^{i}_z+\sum_{i,j} J_{i,j}\hat{\tau}^{i}_z\hat{\tau}^{j}_z+\sum_{i,j,k} G_{i,j,k}\hat{\tau}^{i}_z\hat{\tau}^{j}_z\hat{\tau}^{k}_z+...
\end{equation}
Here, the $\hat{\tau}^{i}_z$ operators are so-called local integrals of motion, pseudo-$\hat{\sigma}_z$ operators that are quasi-local in space.  

Because $\mathcal{H}_{\rm LIOM}$ only involves products of $\hat{\tau}_z$ operators, each $\hat{\tau}^{j}_z$ commutes with the Hamiltonian and thus their occupancies are conserved quantities. Furthermore, as these operators only have support on a few sites, there will be no transport of energy, and an MBL system prepared initially out of equilibrium will never thermalize.

The interacting terms~(e.g., $J_{i,j}$ for two-body interactions) have exponentially decreasing amplitude with the distance between modes~\cite{serbyn2013universal,huse2014phenomenology,chiaro2019growth}. The interactions differentiate MBL from Anderson localization, and cause each local integral of motion $\hat{\tau}_z^i$ to acquire different phases in evolution depending on the occupancy of other $\hat{\tau}_z^j$. For closed systems, this results in a dephasing for any local subsystem, due to the gradual entanglement of these degrees of freedom with others further away. Likewise, starting from a product state in the physical basis, these interactions cause a subsystem $\hat{\rho}_A$ consisting of half the entire system to exhibit a logarithmic growth of entanglement entropy $S(t)=-\text{Tr}[\hat{\rho}_A(t) \operatorname{log}(\hat{\rho}_A(t))]$ in time in the MBL phase~\cite{bardarson2012unbounded}. As a comparison, systems that do not exhibit MBL experience a ballistic growth of entanglement entropy (linear with time). 

\section{Many-body localization in the Hermitian waveguide model}\label{sec:MBLclosed}

In this section, we study the Hermitian component $\mathcal{H}_{\rm hermi}$ of the effective Hamiltonian, looking at transport and entanglement entropy. Specifically, we consider initial states consisting of product states of $n_{\rm exc}$ excitations, $|\psi_{\rm init}\rangle = \prod_{j \in \mathcal{E}_{\rm init}} \hat{\sigma}_{eg}^j |g\rangle^{\otimes N}$  with $\mathcal{E}_{\rm init}$ being the set of indices of the initially excited atoms. Enforcing a fixed number of excitations enables us to study modestly larger system sizes, by throwing out the Hilbert subspace with higher excitation numbers.

We first consider transport, in a system of $N=30$ atoms. In Fig.~\ref{fig:transportclosed}(a), we show the time-dependent excited-state population $\langle \hat{\sigma}^i_{ee}(t)\rangle$ per site, averaged over $\sim 20-100$ disordered configurations, for $n_{\rm exc}=\{1,3,5\}$  roughly equidistantly spaced excitations. In comparison, in Fig.~\ref{fig:transportclosed}(b), we show the evolution for the same initial states, but in an ordered chain of lattice constant $d=2.7\pi/k_\text{1D}$.
In Fig.~\ref{fig:transportclosed}(c) we plot the site-dependent population $\langle \hat{\sigma}^i_{ee}(t_{\rm max})\rangle$ for the final time of the simulation, $t_{\rm max}=100/\Gamma_\text{1D}$, both for the disordered~(blue) and ordered cases~(red). In the ordered system, the population rapidly becomes distributed evenly among all the spins, so that $\langle \hat{\sigma}^i_{ee}(t_{\rm max})\rangle\approx n_{\rm exc}/N$~(dashed black line). (The only exception is for $n_{\rm exc}=1$, where the single excitation evolves according to a well-defined band dispersion relation). In contrast, in the presence of disorder, the system retains a clear memory.

\begin{figure}[h]   
\begin{center}
     \includegraphics[width=0.40\textwidth]{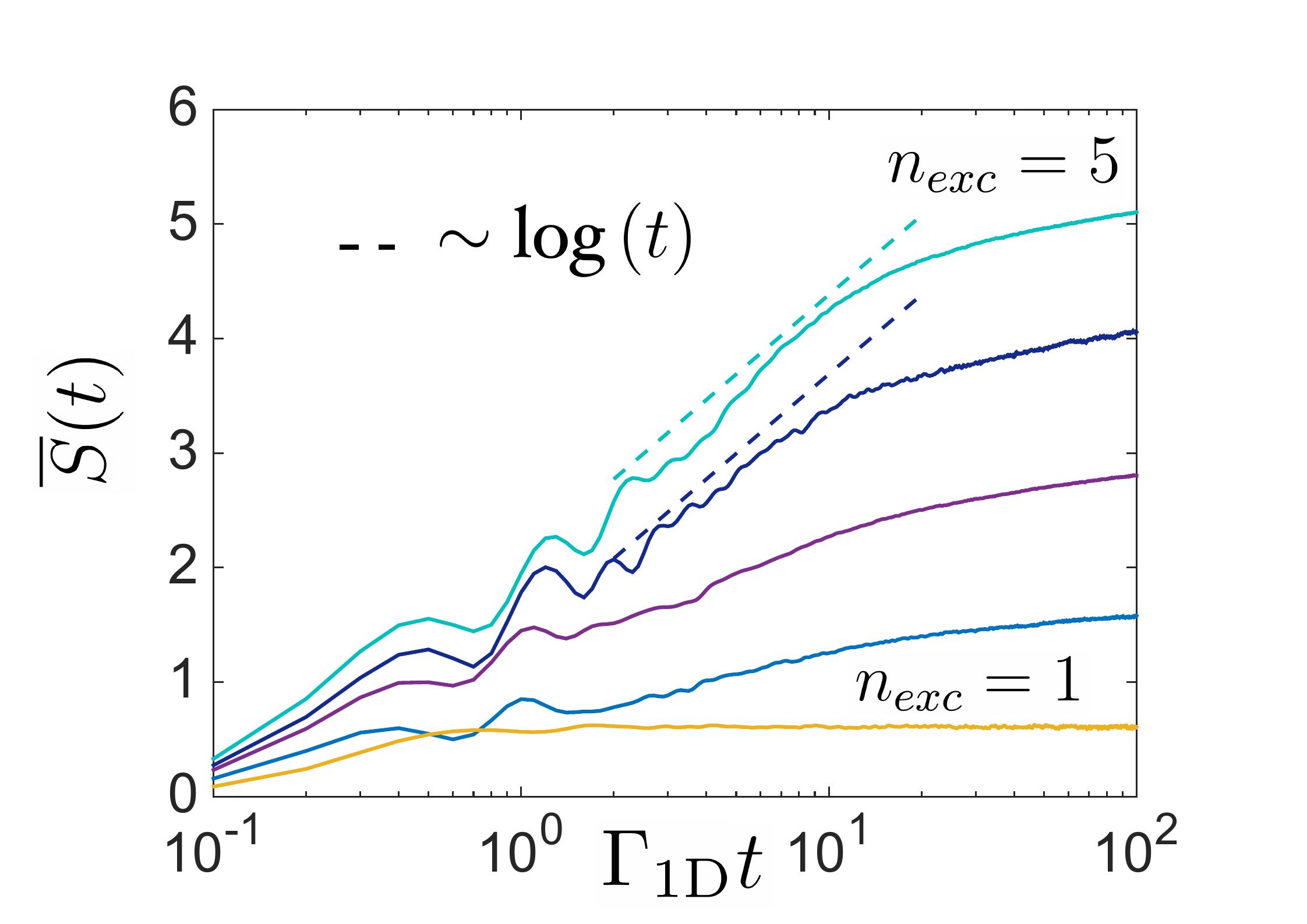} 
             \end{center}
     \caption{ Disorder-averaged half-chain entanglement entropy $\overline{S}(t)$ as a function of time, for different number of atomic excitations initially prepared in a product state, in the Hermitian waveguide. The simulations are for $N=30$ atoms, and the curves from bottom to top correspond to $1 \leq n_{\rm exc} \leq 5$ excitations. Logarithmic scalings~(dashed lines) are shown as a guide to the eye.} \label{fig:EEgrowth}   
\end{figure}

This can be quantified by defining an observable that captures the memory of the initial state:
\begin{equation}
\mathcal{M}(t)=\left(\frac{\sum_{i\in \mathcal{E}_{\rm init}}\langle\hat{\sigma}^{i}_{ee}(t)\rangle}{n_{\rm exc}}-\frac{n_{\rm exc}}{N} \right)/\left(1-\frac{n_{\rm exc}}{N}\right).
\end{equation}

This quantity would be equal to 1 in the case of an initial population distribution conserved through time, and to $0$ in the case of an equally shared population between all atoms. In  Fig.~\ref{fig:transportclosed}(d), we plot the disorder-averaged $\overline{\mathcal{M}}(t)$ and observe a clear memory of the initial state (or, equivalently, an absence of transport at long times) in all the cases with disorder. 

In Fig.~\ref{fig:EEgrowth} we study the evolution of the half chain entanglement entropy $S(t)$ in time, for the same disordered system and initial conditions. A clear region of logarithmic growth is observed when the number of excitations is large enough (up to the largest number $n_{\rm exc}=5$ we can simulate in a system of $N=30$ atoms), before saturating due to the limited system size. Figures~\ref{fig:transportclosed} and~\ref{fig:EEgrowth} provide evidence that the Hermitian system, as defined by $\mathcal{H}_{\rm hermi}$, exhibits a many-body localized phase up to an excitation density of at least $\sim 1/6$.

\subsection{Delocalization transition}

As discussed in Sec.~\ref{sec:description}, based on qualitative arguments, we expect a disordered waveguide QED system to exhibit an MBL phase up to an excitation density of $\rho_{ee}=n_{\rm exc}/N=1/2$ in the thermodynamic limit. However, we also expect that the localization length $N_{\rm loc}(\Delta,\rho_{ee})$ should gradually increase as the excitation density increases, eventually diverging as $\rho_{ee}\rightarrow 1/2$. Thus, for a finite system of moderate size as can be simulated, this would lead to a smooth crossover to delocalization as $n_{\rm exc}$ is increased.

 \begin{figure}[b]   
\begin{center}     
\includegraphics[width=0.80\textwidth]{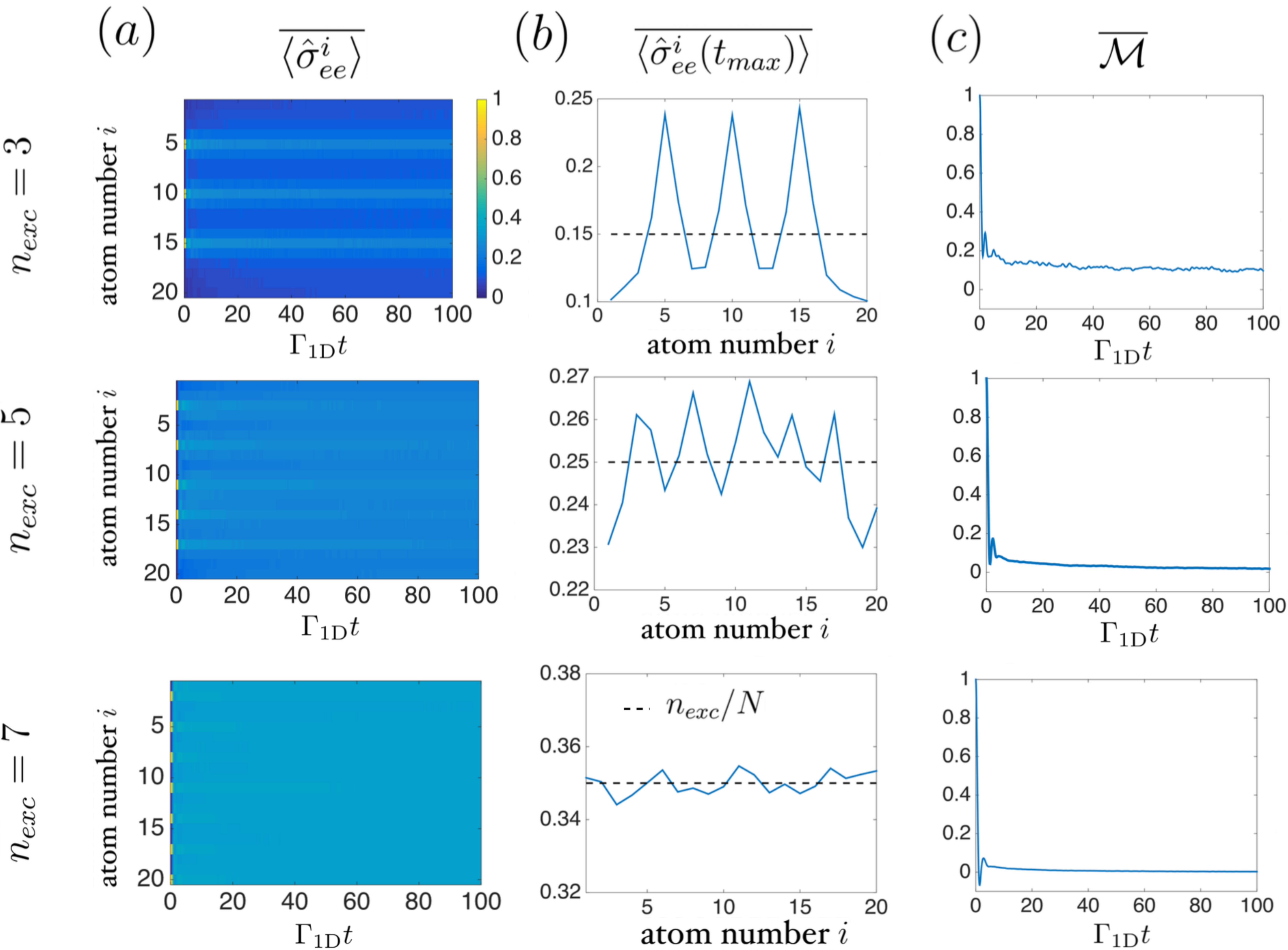} 
             \end{center}
   \caption{ Transport properties of $N=20$ atoms in the Hermitian waveguide, averaged over disorder configurations. Each line corresponds to a different number of excitations~($n_{\rm exc}=3,5,7$), initialized in a product state. (a) Time-dependent, site-resolved excited state population $\langle \hat{\sigma}_{ee}^{i}(t) \rangle$. (b) Site-resolved excited state population at the final simulation time, $t_{\rm max}=100/\Gamma_\text{1D}$. As a guide to the eye, we plot $n_{\rm exc}/N$~(dashed line) to indicate the population if the excitations become equally distributed. (c) Time-dependent memory parameter $\overline{\mathcal{M}}(t)$.}  \label{fig:deloc}   
\end{figure}

\begin{figure}[t]   
\begin{center}     
        \includegraphics[width=0.35\textwidth]{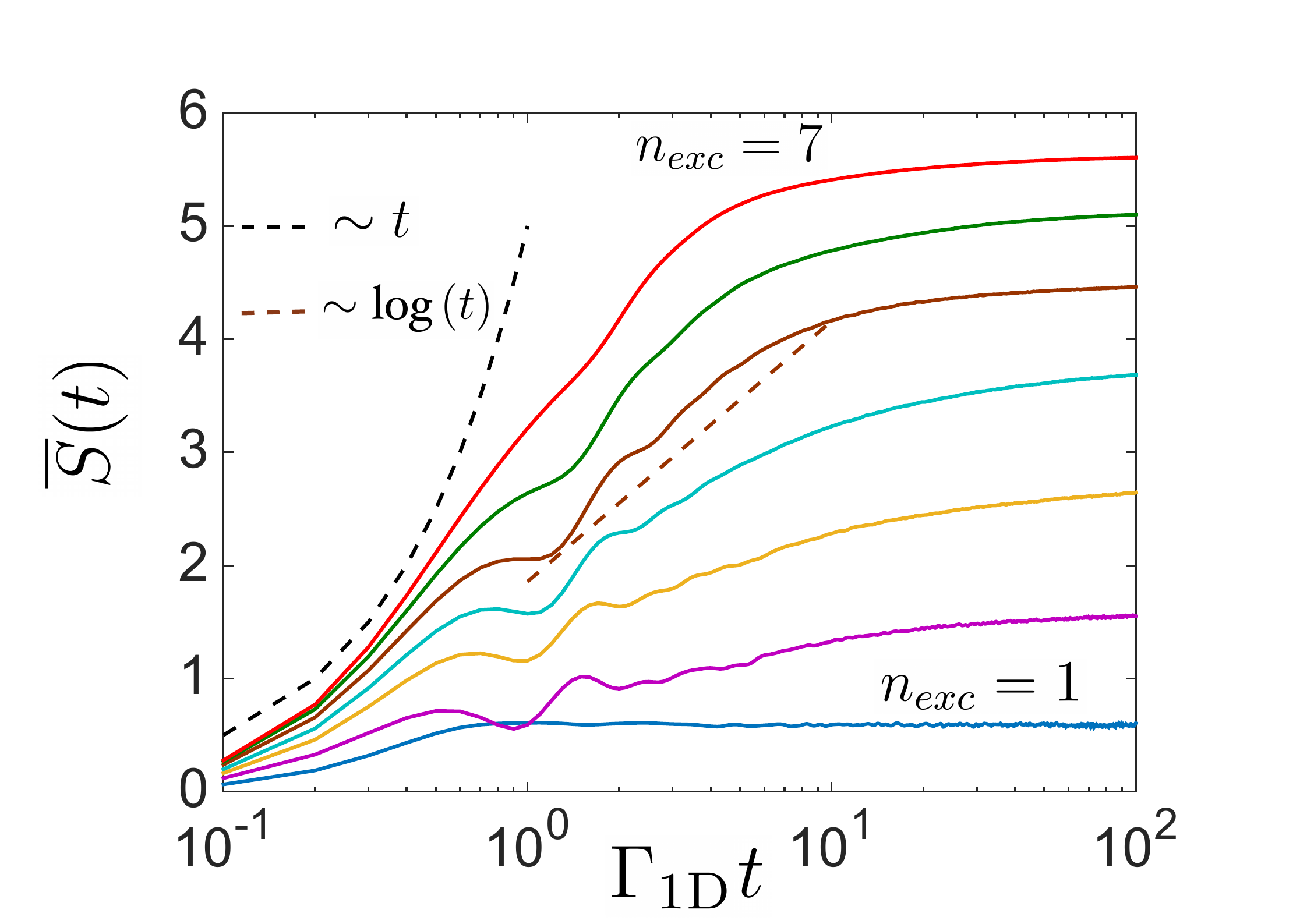} 
             \end{center}
   \caption{Disorder-averaged half-chain entanglement entropy $\overline{S}(t)$ as a function of time, for different number of atomic excitations initially prepared in a product state, in the Hermitian waveguide. The simulations are for $N=20$ atoms, and the curves from bottom to top correspond to $1 \leq n_{\rm exc} \leq 7$ excitations. Linear and logarithmic scalings~(dashed curves) are shown as a guide to the eye. }  \label{fig:EEdeloc}   
\end{figure}

To observe this, we consider a smaller chain of $N=20$, and repeat the same calculations as before for the Hermitian waveguide, but now for $1\leq n_{\rm exc}\leq 7$ to reach a higher excitation density. In Fig.~\ref{fig:deloc}(a) we plot the transport properties for $n_{\rm exc}=3,5$ and $7$ excitations. It can be seen that the memory of the initial state becomes negligible already for $n_{\rm exc}=5$, and essentially vanishes for $n_{\rm exc}=7$, as the initial population is eventually equally shared among all atoms in the waveguide. In Fig.~\ref{fig:EEdeloc} we plot the half-chain entanglement entropy, and see a smooth transition toward ballistic growth as the number of excitations exceeds $n_{\rm exc}\gtrsim 4-5$.

While a delocalization transition that occurs at a lower density of excitations, $n_{\rm exc}/N\sim 1/4$, cannot be ruled out from these numerics, the results are also consistent with our previous hypothesis of a smooth crossover. In particular, assuming that the dynamics have a characteristic bandwidth of $\sim \Gamma_\text{1D}$, one finds a heuristic many-body localization length $N_{\rm loc}(\Delta=\Gamma_\text{1D},\rho_{ee})$ that exceeds the system size of $N=20$ for $n_{\rm exc}\gtrsim 4$.

\section{MBL in the open system}\label{sec:MBLopen}

\subsection{Quantum transport}

We now investigate practical signatures of MBL in the physical, open system. Anticipating that regions within the excitation density-dependent localization length $\sim N_{\rm loc}(\Delta,\rho_{ee})$ of the boundaries are strongly affected by dissipation, we reduce their relative contribution in numerics by going to the open half-waveguide, as described by Eq.~(\ref{eq:masterhalf}), which only has one radiating boundary located to the right.

   \begin{figure}[b]   
\begin{center}
     \includegraphics[width=0.6\textwidth]{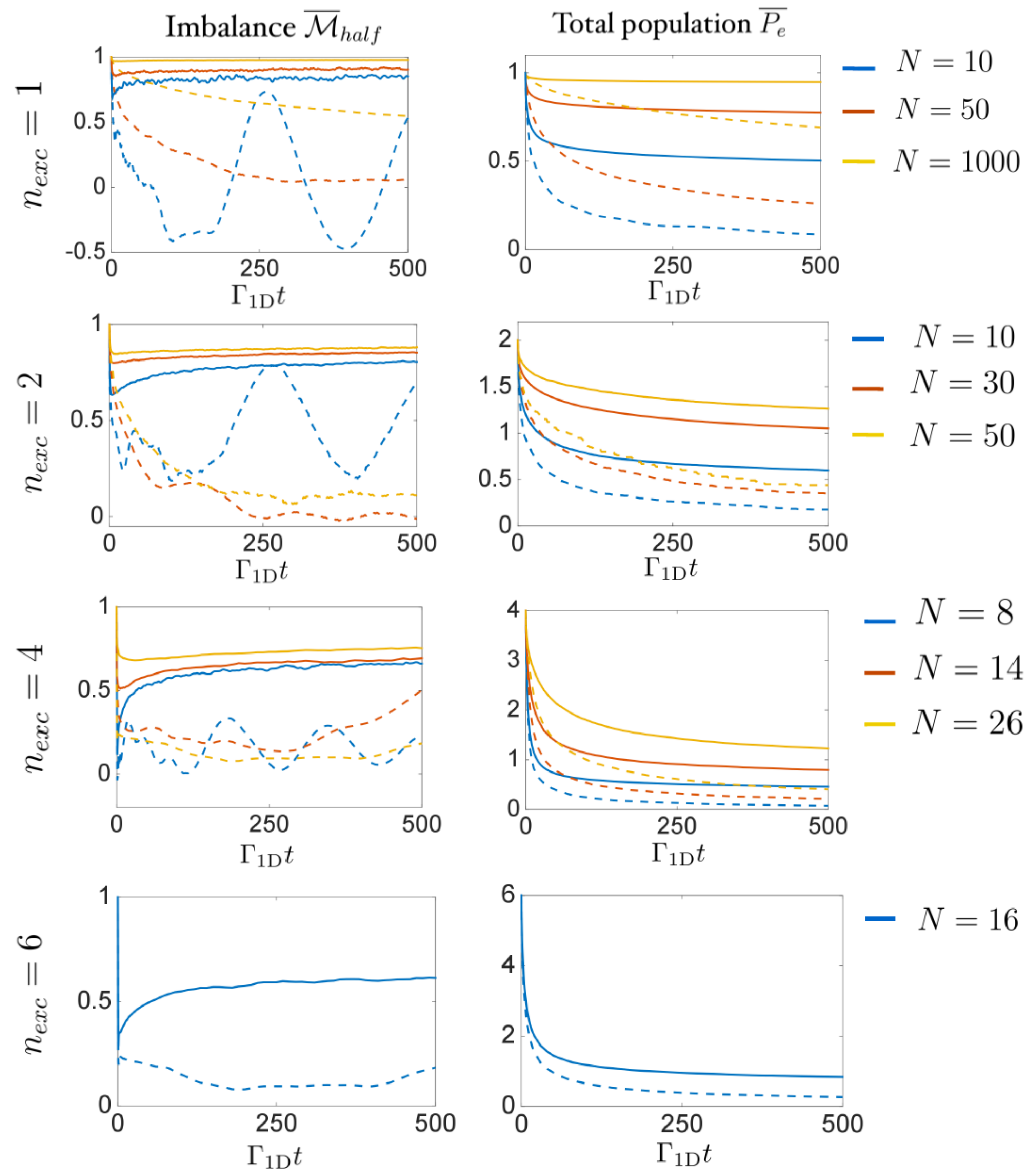} 
             \end{center}
     \caption{
Time-dependent, disorder-averaged population imbalance $\overline{\mathcal{M}}_{\rm half}(t)$~(left panels) and total population $\overline{P_e}(t)$~(right panels) for the open half-waveguide, for $n_{\rm exc}=1,2,4,6$ initial excitations, and for various system sizes $N$~(indicated by different colors). We plot these quantities both for the disordered~(solid curves) and ordered~(dashed) systems.}  \label{fig:transportopen1}   
\end{figure}

We begin by studying transport in the open system. Based on the intuition that the system will resemble a closed system (and thus exhibit stronger features of MBL) provided that excitations remain far from the boundary, we slightly alter the initial conditions compared to Sec.~\ref{sec:MBLclosed}, and take initial states consisting of $n_{\rm exc}$ excitations all located on the left half of the chain. To eliminate any bias originating from a specific choice of location within the left half, we consider equal superpositions of all possible basis states, with each basis state multiplied by a random phase. We have checked, however, that other choices do not affect the final conclusions. Numerically, we directly solve for the master equation dynamics of Eq.~(\ref{eq:masterhalf}), for $n_{\rm exc}\leq 2$, and by a quantum jump approach for higher excitation number, with a minimum of $150$ trajectories for each configuration, and with at least $50$ random configurations in all cases to calculate disorder-averaged observables.

We then define the time-dependent imbalance $\mathcal{M}_{\rm half}(t)$ between the left and right halves,
\begin{equation}
\mathcal{M}_{\rm half}(t)=\frac{1}{P_e(t)}\left(\sum_{i<N/2+1}\langle \hat{\sigma}_{ee}^i (t)\rangle-\sum_{i>N/2}\langle \hat{\sigma}_{ee}^i(t) \rangle\right),
\end{equation}
where $P_e(t)=\sum_i \langle \hat{\sigma}_{ee}^i(t) \rangle$ is the total excited population that now decays in time. In Fig.~\ref{fig:transportopen1}~(left panel) we plot the disorder averaged, time-dependent imbalance for various initial excitation numbers $n_{\rm exc}=1,2,4,6$ and for different system sizes (solid lines). Alongside the imbalance, we also plot $P_e(t)$.

Beginning with a low density of excitations~($n_{\rm exc}=1,2$) in the disordered system, we see that both transport and dissipation are strongly suppressed, with the asymptotic behavior apparently trending toward perfect suppression, $\mathcal{M}_{\rm half}\rightarrow 1$ and $P_e\rightarrow n_{\rm exc}$, as $N\rightarrow \infty$. This confirms the increasingly closed nature of the  system.
Similar trends also are observed for a higher number of excitations, $n_{\rm exc}=4$. Here, however, the localization length $N_{\rm loc}(\Delta=\Gamma_\text{1D},\rho_{ee})$~(again assuming a characteristic bandwidth $\sim\Gamma_\text{1D}$ for dynamics) becomes comparable to the system sizes simulated, and dissipation is no longer negligible. Interestingly, one sees that the imbalance does not monotonically decay, but rather reaches a minimum over a short time scale, while partially recovering at later times. This behavior is more prominent for smaller $N$, due to the higher initial excitation density. Physically, part of the initial population on the left experiences transport and quickly propagates toward the right boundary of the chain (decreasing the imbalance), where it can be dissipated. The lowering of density on the left half subsequently suppresses the transport, and the system \textit{dynamically} enters into an MBL phase, while the imbalance partially recovers as the remaining excitations on the right half then slowly but preferentially dissipate away. Given this observation, we are then motivated to investigate even higher initial densities ($n_{\rm exc}=6$ and $N=16$), where now the localization length $N_{\rm loc}(\Delta=\Gamma_\text{1D},\rho_{ee}=3/8)\approx 65 \gg N$ far exceeds the system size. Despite clearly starting in a transport and dissipation-allowed phase, the same dynamical behavior is clearly and even more prominently observed.

We now return to the hypothesized phase diagram of Fig.~\ref{fig:schematic}(b), which we initially considered to be for a system in the thermodynamic limit. Considering that any physical system will consist of finite atom number, we propose that its dynamical behavior, if it begins in the delocalized phase, will consist of transport-facilitated dissipation, until it reaches an MBL phase, as illustrated by the arrow showing evolution in time. Furthermore, this dissipation process should be quite rapid and dominate the initial dynamics, due to the lack of subradiant states~(with decay rate $\ll\Gamma_{\rm 1D}$) for excitation densities $n_{\rm exc}/N\gtrsim 1/2$, and by the fact that the \textit{average} dissipation rate of eigenstates of the non-Hermitian Hamiltonian for such high excitation densities must grow extensively with system size, $\sim N\Gamma_{{\rm 1D}}$.   

Finally, for comparison, we plot with dashed curves the same observables for all the system sizes and number of excitations previously considered, but for an ordered system. It should be noted that the total population in the ordered system also decays slowly for long times, albeit faster than the disordered case. This can be attributed to the large number of single and multi-excitation subradiant states in a 1D waveguide at low excitation density~\cite{albrecht2019subradiant}. In the case of a full waveguide~(dissipation at both boundaries) and sufficient number of excitations, it has previously been shown that this leads to a power-law decay for ordered systems at long times~\cite{henriet_clock}.

\subsection{Quantum revivals}

\begin{figure}[b]   
\begin{center}
     \includegraphics[width=0.90\textwidth]{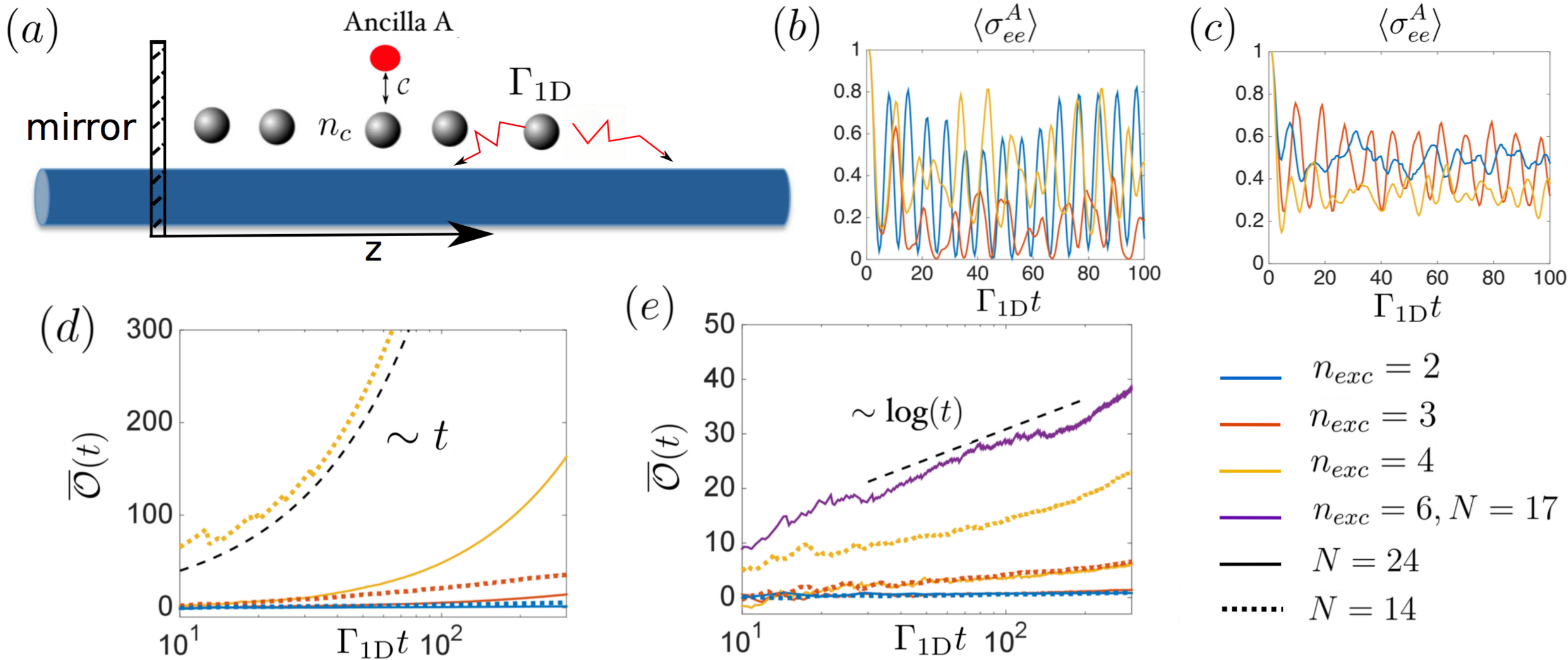} 
             \end{center}
     \caption{ (a) Scheme of the quantum revival setup to measure the interacting degrees of freedom in the open half-waveguide. An additional ancilla $A$ is introduced, which does not couple or radiate into the waveguide, but couples to and can exchange excitations with the waveguide atom $n_c$. The time-dependent excited population of the ancilla is monitored to record the number of revivals $N_{\rm rev}$ within a certain time window $[0,t]$. (b,c) Ancilla population versus time, for three typical disorder realizations for a Hermitian half-waveguide. Here, initially the ancilla is excited, while the waveguide atoms contain no excitations (b) and $2$ excitations located at positions $\vert e_1,e_4\rangle$ (c). The total system size (ancilla $+$ atoms) is $N=14$. Disorder-averaged $\mathcal{O}(t)$ for both Hermitian (d) and open (e) disordered systems of size $N=14$ (dotted curves), $N=17$ (purple curve)  $N=24$ (solid curves) with varying number of excitations ($n_{\rm exc}=2,3,4,6$) indicated by different colors. As a guide to the eye, we show a linear evolution in (d) and a logarithmic evolution in (e) in black dashed curves.} \label{fig:QRfig1}    
\end{figure}

As entanglement entropy is difficult to measure in a closed system of large size, and has no immediate analogue in an open system, capturing some essence of the slow entanglement entropy growth is generally one of the most challenging aspects of experimentally studying MBL~\cite{lukin2019probing}. When the local integral of motion closely resembles the physical spin, it has been theoretically proposed that the slow growth of entanglement can be probed by generalized spin echo techniques~\cite{serbyn2014interferometric}. In our system, however, we find that this technique is ineffective, as localized modes have a non-negligible extent over at least a few sites (see the single-excitation dynamics in Fig.~\ref{fig:transportclosed}, for example). Physically, this is because suppressing the emission of light by one excited atom arises from the excitation of at least the neighboring atoms and their destructive interference in radiation. 

We thus adopt an alternative proposal by Vasseur et al. \cite{vasseur2015quantum}, which is readily implementable within waveguide QED. In particular, we consider the addition of an extra ancilla atom $A$ that does not directly couple or radiate into the waveguide, but is allowed to coherently exchange excitations with the waveguide atom with index $n_c$~[see Fig.~\ref{fig:QRfig1}(a)]. The waveguide atoms thus serve as an exotic bath for the ancilla. The corresponding Hamiltonian for this system~(studied for the half-waveguide) is
\begin{equation}
H=\mathcal{H}_{\rm half}+\mathcal{H}_{c},
\end{equation}
with the non-Hermitian Hamiltonian $\mathcal{H}_{\rm half}$ given in Eq.~(\ref{eq:Hhalf}) and $\mathcal{H}_c=\mathcal{C}\left(\hat{\sigma}_{eg}^A\hat{\sigma}_{ge}^{n_c}+h.c.\right)$ coupling the ancilla to atom $n_c$ in the chain. In the following we use $\mathcal{C}=\Gamma_\text{1D}/2$ and we couple the ancilla to the atom number $n_c=3$ in the chain of $N-1$ atoms.

If the system was governed purely by $\mathcal{H}_c$, an ancilla initially excited would indefinitely and reversibly exchange its excitation with atom $n_c$. In the Anderson localized regime, where the waveguide interactions are turned on but all waveguide atoms are initially unexcited, the ancilla couples to a few localized modes (those whose support contain atom $n_c$). This leads to a more complex structure of collapse and revival dynamics in the ancilla population over time, as illustrated in Fig.~\ref{fig:QRfig1}(b) (calculated here for the Hermitian half-waveguide, with no dissipation). Although the details of the revivals depend on the disorder configuration, since the number of localized modes to which the ancilla couples is finite and fixed, the revivals would continue with the same frequency and amplitude indefinitely in the absence of dissipation, and for an increasingly long time with increasing system size in the presence of dissipation. In contrast, in the MBL phase where multiple waveguide atoms are initially excited:
\begin{equation}
\vert \psi_{\rm init} \rangle=    \vert e_{A}\rangle\otimes \vert \psi_{\rm wg}\rangle,
\end{equation}
with $\vert \psi_{\rm wg}\rangle$ containing $n_{\rm exc}-1$ excitations,  
interactions between excitations effectively cause the ancilla to couple to an increasing number of degrees of freedom in time, which makes revivals less likely. This is illustrated in Fig.~\ref{fig:QRfig1}(c), for the case of $n_{\rm exc}-1=2$  excitations initialized in the Hermitian half-waveguide as well, at positions $\vert e_1,e_4\rangle$.

For closed systems, the revival rate should approximately be inversely proportional to the effective size of the Hilbert space to which the ancilla is coupled. We can thus numerically study the revival rate $R(t)=N_{\rm rev}(t)/t$~(see Appendix B for how the number of revivals $N_{\rm rev}(t)$ in the time window $[0,t]$ is determined). To do this, we again directly solve the master equation for low excitation numbers $n_{\rm exc}=1,2$, and by the quantum jump formalism for higher numbers (with at least $500$ trajectories for each configuration). We perform averages over at least 50 configurations for each choice of $n_{\rm exc},N$, and the initial waveguide state is chosen to be in an equal superposition of all basis states with $n_{\rm exc}-1$ excitations, multiplied by random phases.

To quantify the evolution of the average number of degrees of freedom $\overline{\mathcal{N}}(t)$ to which the ancilla is coupled by interactions, we take the ansatz that 
\begin{equation}
\overline{R}(t)\approx \frac{\alpha}{\overline{\mathcal{N}}(t)},
\end{equation}
where $\alpha$ is a proportionality constant. We further split $\overline{\mathcal{N}}(t)=\overline{\mathcal{N}}_0(t)+\overline{\mathcal{N}}_{\rm int}(t)$ with $\overline{\mathcal{N}}_0(t)$ the average number of degrees of freedom that interacts with the ancilla without interactions inside the waveguide and $\overline{\mathcal{N}}_{\rm int}(t)$ the average number of degrees of freedom that comes from interactions between the different qubits. $\overline{\mathcal{N}}_{\rm int}(t)$ should grow logarithmically in time for closed, large MBL systems and linearly in time for delocalized systems, while $\overline{\mathcal{N}_{0}}(t)$ saturates to a constant value that depends on the localization length $N_{\rm loc}$.

Thus, using the unloaded waveguide as a reference to obtain $\overline{R}_0(t)$, we can extract from the revival rates the average number of interacting degrees of freedom (up to a proportionality factor), by considering the quantity
\begin{equation}
\overline{\mathcal{O}}(t)\equiv 1/\overline{R}(t)-1/\overline{R}_0(t)\approx \frac{ \overline{\mathcal{N}}_{\rm int}(t)}{\alpha}.
\label{eq:Extract}
\end{equation}

We plot $\overline{\mathcal{O}}(t)$ for different initial excitation numbers $n_{\rm exc}=2,3,4,6$ and atom number $N=14,17,24$ in Fig.~\ref{fig:QRfig1}, both for the Hermitian (d) and open systems (e) up to $\Gamma_{\rm 1D}t=300$. Starting with the Hermitian half-waveguide system, one clearly observes a transition from logarithmic to linear growth of $\overline{\mathcal{O}}(t)$ as the system goes from a MBL to a delocalized phase when the density of excitations increases. Interestingly, for the open system, we observe a slow logarithmic growth of $\overline{\mathcal{O}}(t)$ in all the cases considered over this time range. We again attribute this to a dynamical transition, where transport results in rapid dissipation of large excitation densities, until the waveguide atoms reach an MBL phase. This transition occurs faster than the characteristic interaction rate of the ancilla with the bath~(taken to be comparable to the single atom-waveguide coupling strength $\Gamma_{\rm 1D}$), resulting in suppression of quantum revivals reminiscent of a closed MBL system. Separately, we have seen that for sufficiently long times $\Gamma_{\rm 1D} t \gtrsim 500$, a deviation from logarithmic scaling can be observed, due to the very slow dissipation of remaining excitations left in the system.

\section{Conclusion}\label{sec:Conclusion}

In summary, we have proposed and presented numerical evidence that a system of disordered atoms coupled to a waveguide exhibits an MBL phase, provided that the density of atomic excitations is less than $1/2$. Compared with many other MBL systems already studied, this system has a number of interesting features. In particular, the system contains two types of particles~(atomic spins, and a continuum of photons), and furthermore, the continuum nature of the photon modes intrinsically causes the atomic dynamics to appear open, with a dissipation strength that a priori is equal to the coherent interaction strength, as seen in Eq.~(\ref{eq:master}). Thus, beyond typical MBL systems where dissipation is simply an unwanted effect, our system presents interesting opportunities to study the apparent transition toward Hermitianity~\cite{hamazaki2019non} when the system is localized, and a dynamical transition toward MBL induced by loss.

State-of-the-art waveguide QED systems involving superconducting qubits coupled to transmission lines should already be capable of studying the proposed phenomena. In particular, it has been demonstrated that systems consisting of at least $N=7$ controllable, individually measurable, and \textit{identical} qubits can be realized~\cite{mirhosseini2019cavity}, with ratios of atom-waveguide interaction strengths to additional, unwanted dissipation of $\Gamma_\text{1D}/\Gamma'\sim 10^2$-$10^3$~\cite{mirhosseini2019cavity,hamann2018nonreciprocity}  that enable the predicted dynamics to be observed before additional effects set in. 
Furthermore, since our proposed scheme relies on disorder, the use of identical qubits is not necessary, and up to $N=72$ qubits have been realized in such an instance in similar systems~\cite{fitzpatrick2017observation}. Coupling between superconducting qubits and microwave photons can also be realized in two dimensions~\cite{kollar2019hyperbolic}, thus offering promising opportunities to investigate MBL, including in regimes beyond what can be studied directly with numerics. We note that superconducting qubit systems are already being used to investigate MBL~\cite{roushan2017spectroscopic,xu2018emulating,chiaro2019growth}, albeit in regimes where photons are not a central degree of freedom. Moreover, although still in their infancy, systems consisting of atoms coupled to photonic crystal waveguides can also potentially reach the desired combinations of large atom-waveguide interaction strengths~\cite{hung2013trapped,zang2016interaction,goban2015superradiance,hood2016atom} and large atom number to investigate MBL.

Beyond our initial theoretical investigations, our work also stimulates other theoretical questions to explore. First, while we have focused solely on position disorder~(and full disorder in the numerics, to minimize the localization length), our qualitative arguments about the existence of MBL seem quite general. We thus envision future efforts to confirm a thermodynamic phase diagram similar to Fig.~\ref{fig:schematic}(b), for arbitrary amounts and types of disorder~(e.g., in resonance frequencies). Furthermore, thus far, we have reduced the complexity of our system by integrating out the photons and focusing on the atomic ``spin'' degrees of freedom. While this is an excellent approximation near resonance, it would be interesting to more fully explore the system from the photonic standpoint. For example, we anticipate that the MBL phase is reflected in interesting quantum correlations of light, either generated through the excited atoms themselves, or explicitly via quantum transport by sending in optical pulses. Including the photons might also provide an avenue to develop diagrammatic techniques~\cite{lang2020nonequilibrium,cherroret2016self,wellens} to understand MBL and the delocalization transition, and in a way that is not as limited by system size as with pure numerics. Finally, while we have considered the most basic continuum of photon modes here, consisting of a linear dispersion and infinite bandwidth, current waveguide QED systems also offer excellent potential for dispersion engineering, such as through the introduction of band edges and gaps~\cite{joannopoulos1997photonic} or even its global shape~\cite{kollar2019hyperbolic,nguyen2018symmetry}, and other features such as realizing some degree of chirality in interactions~\cite{petersen2014chiral}. These can dramatically alter the nature and the range of the photon-mediated interactions, and result in non-trivial boundaries between MBL and delocalized phases.

\section{Acknowledgements}
The authors would like to acknowledge I. Aleiner, N. Cherroret, D. Delande, J.J Greffet, A. Goetschy and A. Albrecht for fruitful discussions. DEC acknowledges support from the European Union’s Horizon 2020 research and innovation programme, under European Research Council grant agreement No 639643 (FOQAL) and FET-Open grant agreement No 899275 (DAALI); the Government of Spain, through the Europa Excelencia program (EUR2020-112155, project ENHANCE), Severo Ochoa program CEX2019-000910-S, and Plan Nacional Grant PGC2018-096844-B-I00); Generalitat de Catalunya through the CERCA program, Fundació Privada Cellex, Fundació Mir-Puig, and Secretaria d'Universitats i Recerca del Departament d'Empresa i Coneixement de la Generalitat de Catalunya, co-funded by the European Union Regional Development Fund within the ERDF Operational Program of Catalunya (project QuantumCat, ref. 001-P-001644). AAG acknowledges support from the National Science Foundation QII-TAQS (Grant No. 1936359).

\appendix\section{\\Estimation of the energy density dependent localization length}

In this appendix, we derive the energy density dependent localization length of the 1D disordered and open waveguide.
To do so, let us consider a single two level atom driven by an incoming coherent state of Rabi frequency $\Omega$ and frequency $\omega$, and a detuning of $\Delta=\omega-\omega_0$ relative to the atomic transition frequency. In the rotating frame, the interaction Hamiltonian between this field and a two level atom is
\begin{equation}
H=-\Delta\hat{\sigma}_{ee}+\Omega (\hat{\sigma}_{eg}+\hat{\sigma}_{ge}).
\end{equation}

 \begin{figure}[b]   
\begin{center}     
\includegraphics[width=0.80\textwidth]{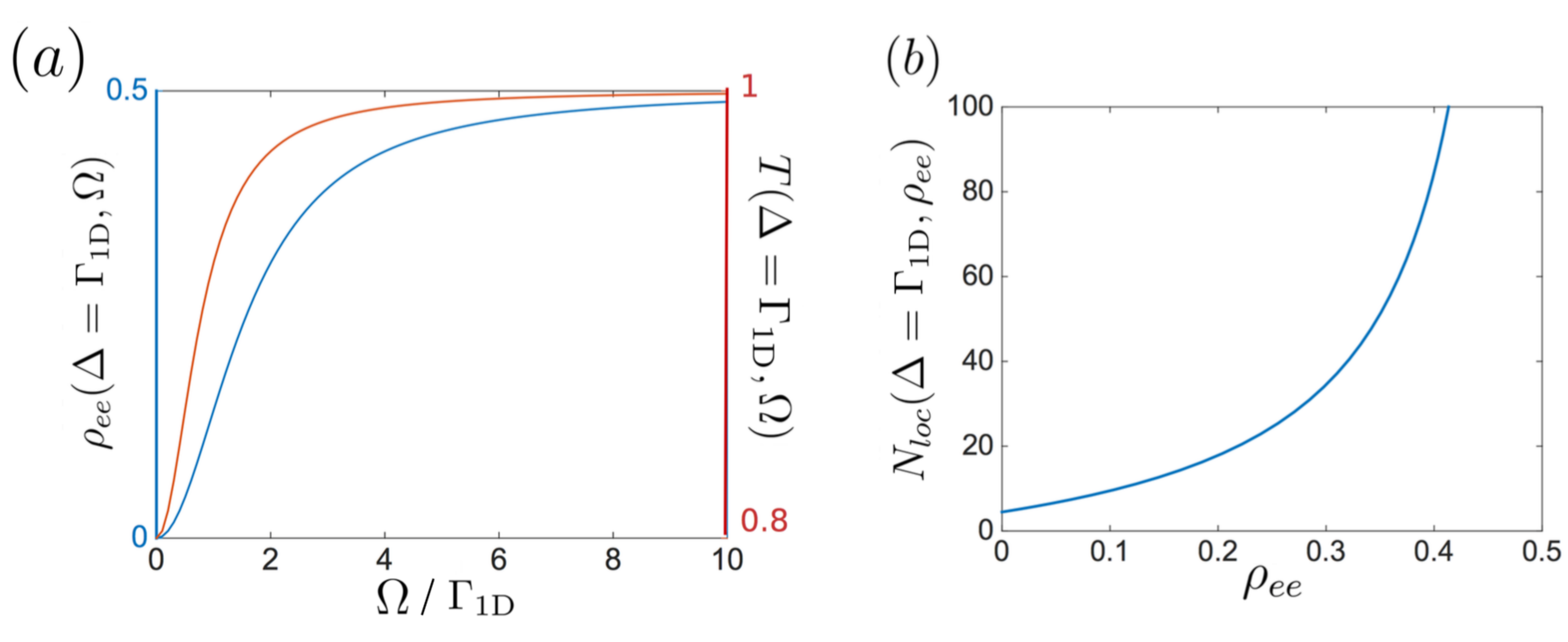} 
             \end{center}
   \caption{
   (a) Excited-state population $\rho_{ee}$ and transmittance $T$ of a single atom as a function of driving Rabi amplitude $\Omega/\Gamma_{\rm 1D}$, and at a detuning $\Delta=\Gamma_{\rm 1D}$. (b) The estimated MBL localization length $N_{\rm loc}(\Delta=\Gamma_{\rm 1D},\rho_{ee})$ as a function of atomic population.}  \label{fig:SI}   
\end{figure}

Inserting this Hamiltonian into the master equation for the two level atom,
\begin{equation}
\frac{\mathrm d \rho}{\mathrm d t}=-i\left[H,\rho\right]+\frac{\Gamma_\text{1D}}{2}\mathcal{L}[\rho]
\end{equation}
with $\mathcal{L}[\rho]=2\hat{\sigma}_{ge}\rho\hat{\sigma}_{eg}-\hat{\sigma}_{ee}\rho-\rho\hat{\sigma}_{ee}$, one obtains the optical Bloch equations whose steady state solutions are:
\begin{equation}
\rho_{ee}=\frac{\Omega^2}{4\Delta^2+\Gamma_\text{1D}^2+2\Omega^2}\label{eq:rhoee}
\end{equation}
and 
\begin{equation}
\rho_{eg}=i\Omega\frac{\Gamma_\text{1D}+2i\Delta}{4\Delta^2+\Gamma_\text{1D}^2+2\Omega^2}.
\end{equation}
The input-output equation for a waveguide~\cite{Chang_2012} allows one to express the transmitted field as
\begin{equation}
 \hat{E}=\hat{E}_{\rm in}+i\sqrt{\frac{\Gamma_\text{1D}}{2}}\hat{\sigma}_{ge}
\end{equation}
where $\hat{E}_{\rm in}$ represents the (coherent state) input field. This leads to the expression of the transmittance $T=\langle \hat{E}^{\dagger}\hat{E}\rangle/\langle \hat{E}^{\dagger}_{\rm in }\hat{E}_{\rm in }\rangle$:
\begin{equation}
T=\frac{4\Delta^2+8\Omega^2}{\Gamma_\text{1D}^2+4\Delta^2+8\Omega^2}.\label{eq:Tsaturation}
\end{equation}
From this single-atom transmittance, we can estimate the saturation and frequency dependent localization length using
\begin{equation}
N_{\rm loc}(\Delta,\rho_{ee})=1/\vert \operatorname{log}(T) \vert\label{eq:localizationlength},
\end{equation}
with $N_{\rm loc}$ being the localization length in units of number of atoms. In the equation above, the excited-state population $\rho_{ee}$ parametrically depends on the Rabi frequency via Eq.~(\ref{eq:rhoee}).

 In a waveguide QED system, one can estimate that the characteristic bandwidth of MBL dynamics is given by $\Gamma_{\rm 1D}$. One can then evaluate the excited-state population $\rho_{ee}$ and transmittance $T$ at a detuning $\Delta=\Gamma_{\rm 1D}$~(illustrated in Fig.~\ref{fig:SI}(a), as a function of $\Omega/\Gamma_{\rm 1D}$), and subsequently calculate the excited-state population-dependent MBL localization length $\sim N_{\rm loc}(\Delta=\Gamma_{\rm 1D},\rho_{ee})$, which we plot in Fig.~\ref{fig:SI}(b). Note that the localization length exceeds the maximum system sizes we can numerically study $N\sim 20$-$30$ for excitation densities of $n_{\rm exc}/N\sim 0.3$.

\section{\\Revival counting}
In the quantum revival simulations, one has to count the revival rates of the ancilla in order to extract the number of interacting degrees of freedom in the waveguide.
For each realization of the disorder, we obtain the population of the ancilla $\langle \hat{\sigma}_{ee}^A (t)\rangle$ as a function of time from which we extract all the times $t_n$ corresponding to an extremum of $\langle \hat{\sigma}_{ee}^A (t)\rangle$. Our algorithm is designed to prevent local maxima with arbitrarily small visibility from qualifying as true revivals. To this end, for each local maximum, we compute $Q=\frac{\langle \hat{\sigma}_{ee}^A(t_n)\rangle-\langle \hat{\sigma}_{ee}^A\rangle_{{\rm min},n-1}}{\langle \hat{\sigma}_{ee}^A\rangle_{{\rm min},n-1}}$, which  compares the value at the position of the maximum with  the minimum value of the population of the ancilla, during the time since the last revival maximum. We count the local maximum at $t_n$ as a true revival if $Q$ exceeds a prescribed value $Q_{\rm min}$, and if the population of the ancilla $\langle \hat{\sigma}_{ee}^A(t_n)\rangle$ exceeds $0.25$. For each disorder realization, we then define $N_{\rm rev}(t)$ as the cumulative number of revivals of the population in the time window $[0,t]$. Then we average $N_{\rm rev}(t)$ over the disorder in order to obtain $\overline{R}(t)=\overline{N_{\rm rev}}(t)/t$ (note that the details of the revival dynamics vary for each configuration, so one cannot obtain $\overline{R}(t)$ from the disorder-averaged population of the ancilla $\overline{\langle \hat{\sigma}_{ee}^A (t)\rangle}$). All the curves presented in this work have been obtained taking $Q_{\rm min}=0.4$. 

\newpage

\bibliographystyle{naturemag} 

\bibliography{biblioMBL}

\end{document}